\documentclass[fleqn,usenatbib]{mnras}

\usepackage[T1]{fontenc}
\usepackage{physics}
\usepackage{tabularx}
\DeclareRobustCommand{\VAN}[3]{#2}
\let\VANthebibliography\thebibliography
\def\thebibliography{\DeclareRobustCommand{\VAN}[3]{##3}\VANthebibliography}
\defcitealias{sadeh_non-thermal_2023}{Paper 1}

\usepackage{graphicx}	
\usepackage{amsmath}	
\usepackage{amssymb}	
\usepackage{subcaption}
\usepackage{newtxtext,newtxmath}

\title[Non-thermal emission of BNS merger]{Non-thermal emission from mildly relativistic dynamical ejecta of neutron star mergers: spectrum and sky image}

\author[G. Sadeh et al.]{
Gilad Sadeh\thanks{E-mail: gilad.sade@weizmann.ac.il}, Noya Linder, and Eli Waxman
\\
\it{Dept. of Particle Phys. \& Astrophys., Weizmann Institute of Science, Rehovot 76100, Israel}
}

\date{Accepted XXX. Received YYY; in original form ZZZ}

\pubyear{2024}

\begin{document}
\label{firstpage}
\pagerange{\pageref{firstpage}--\pageref{lastpage}}
\maketitle

\begin{abstract}
Binary neutron star mergers are expected to produce fast dynamical ejecta, with mildly relativistic velocities extending to $\beta= v/ c > 0.6$. In a preceding paper, we derived an analytic description of the time-dependent radio to X-ray synchrotron flux produced by collisionless shocks driven by such fast ejecta into the interstellar medium, for spherical ejecta with broken power-law mass (or energy) distributions, $M(>\gamma\beta)\propto(\gamma\beta)^{-s}$ with $s=s_{\rm KN}$ at $\gamma\beta<\gamma_0\beta_0$ and $s=s_{\rm ft}$ at $\gamma\beta>\gamma_0\beta_0$ (where $\gamma$ is the Lorentz factor). Here, we extend our analysis and provide analytic expressions for the self-absorption frequency, the cooling frequency, and the observed angular size of the emitting region (which appears as a ring in the sky). For parameter values characteristic of merger calculation results - a "shallow" mass distribution, $1<s_{\rm KN}<3$, for the bulk of the ejecta (at $\gamma\beta\approx 0.2$), and a steep, $s_{\rm ft}>5$, "fast tail" mass distribution – the analytic results reproduce well (to tens of percent accuracy) the results of detailed numeric calculations,
a significant improvement over earlier order-of-magnitude estimates (based on extrapolations of results valid for $\gamma\beta\ll 1$).
\end{abstract}




\section{Introduction}
Mergers of binary neutron stars (BNS) are expected to produce mildly relativistic ($\gamma\beta>0.1$) ejecta that propagate into the interstellar medium (ISM)  \citep[see][for reviews]{fernandez_electromagnetic_2016,radice_dynamics_2020}. The structure, velocity, and geometry of these ejecta are currently explored by numerical relativity (NR) simulations of BNS mergers. 
Those simulations suggest that a small fraction ($10^{-7}M_\odot$ to $10^{-4}M_\odot$) of the ejecta mass (the "fast tail") may reach relativistic velocities ($v>0.6c$), while the bulk of the ejecta ($10^{-3}M_\odot$ to $5\times10^{-2}M_\odot$) is expected to attain velocities of $\sim 0.1-0.3c$ \citep{radice_binary_2018,nedora_dynamical_2021,nedora_numerical_2021,fujibayashi_comprehensive_2023,hajela_evidence_2022,radice_new_2022,rosswog_heavy_2024}.
Radioactive heating of the bulk of the ejecta is expected to produce IR-UV "Kilonova" (KN) emission \citep{li_transient_1998}, while the fast tail is predicted to generate a non-thermal radio to X-ray flux on a time scale of years by synchrotron emission from collisionless shocks driven by the fast ejecta into the interstellar medium \citep{nakar_radio_2011,sadeh_non-thermal_2023}. \cite{ghosh_search_2024} found no evidence of late radio emission in a follow-up campaign of five short gamma-ray bursts, which are considered to originate in compact objects mergers. The expected peak flux of such a late-time component following BNS mergers at the distances examined in \cite{ghosh_search_2024}, >$500$Mpc, is $\sim 0.5\mu$Jy \citep{sadeh_non-thermal_2023}, well below the Very Long Array (VLA) sensitivity.

The mass and velocity profile of the fast tail depends strongly on the parameters of the binary system and on the EoS. Depending on parameter values, the ejecta mass at $\beta> 0.6$ varies, for example, between $10^{-7}M_\odot$ and $10^{-4}M_\odot$, and the maximal velocity varies from $\gamma\beta\approx0.6$ to $\gamma\beta> 3$. This implies that measurements of the non-thermal emission driven by the fast ejecta may provide stringent constraints on model parameters. Two points should, however, be noted in this context. First, a reliable determination of constraints would require a significant reduction of the current large numerical uncertainty in the numerically calculated fast tail parameters, as reflected by the large variations of results between different simulations \citep[e.g.][]{radice_binary_2018, dean_resolving_2021,nedora_dynamical_2021}. 

Second, if BNS mergers are the sources of (short) gamma-ray bursts \citep[see][for reviews]{meszaros_theories_2002,piran_physics_2004,nakar_short-hard_2007}, then they are expected to launch highly relativistic jets that, depending on the jet energy, opening and observing angles, may dominate the non-thermal emission driven by the "dynamical" (shock and tidally driven) ejecta obtained in NR BNS merger calculations. In the case of the observed non-thermal emission of GW170817 \citep{hallinan_radio_2017,troja_x-ray_2017,lyman_optical_2018}, the swift decay \citep{troja_year_2019, fong_optical_2019, lamb_optical_2019, hajela_two_2019, troja_thousand_2020,  nakar_electromagnetic_2020,makhathini_panchromatic_2021,balasubramanian_continued_2021,balasubramanian_gw170817_2022} and the superluminal motion of the radio centroid suggest a relativistic jet as the driver of the observed non-thermal emission \citep[noting that a $\gamma\sim5$ jet may account for the observations, hence a much higher Lornetz factor cannot be directly inferred,][]{mooley_superluminal_2018,mooley_optical_2022}. Generally, due to the jet's lower mass and higher Lorentz factor, the late-time emission is expected to be dominated by the dynamical ejecta. The interpretation of future observations will require disentangling the two components.

In a preceding paper, \citet{sadeh_non-thermal_2023} (hereafter \citetalias{sadeh_non-thermal_2023}), we considered the non-thermal radio to X-ray synchrotron emission produced by collisionless shocks driven by mildly relativistic ejecta expanding into an interstellar medium (ISM) with a uniform (number) density $n$. Similar to other earlier papers \citep{nakar_radio_2011,kathirgamaraju_em_2019,nedora_synthetic_2023}, we did not provide an accurate description of synchrotron self-absorption, which is expected to affect the emission at low radio frequencies, and of the cooling of electrons on time-scales shorter than the expansion time-scale, that is expected to be important for electrons emitting at hard X-ray bands (and beyond). The analytic expressions we provided hold, therefore, in the frequency range $\nu_a<\nu<\nu_c$, where $\nu_a$ is the self-absorption frequency (below which absorption is significant), and $\nu_c$ is the synchrotron emission frequency of electrons with cooling time comparable to the expansion time. Note that the frequency of synchrotron radiation emitted by the lowest energy electrons, $\nu_m$, is typically much smaller than $\nu_a$ (e.g. Eq. (15) of \citetalias{sadeh_non-thermal_2023}). 

In this paper, we extend our analysis and provide analytic expressions for $\nu_a$, $\nu_c$, and for the observed angular size of the emitting region (which appears as a ring in the sky), which are calibrated using the results of numeric calculations. Observational identification of $\nu_a$ (in radio) and $\nu_c$ (in X-rays), and measurements of the angular size of the source (which may be possible with the Very Long Baseline Interferometry (VLBI) for sources at distances of $\sim100$Mpc, as the size is expected to reach $\approx 10$~pc on years time scale) will provide essential constraints on model parameters. We provide approximations that are valid at times earlier than the peak time, $t_\text{peak}$ at which the non-thermal emission at $\nu_a<\nu<\nu_c$ peaks, which is of the order of a few years.

Similar to \citetalias{sadeh_non-thermal_2023}, we consider spherical ejecta with a broken power-law dependence of mass on momentum,
\begin{equation}
\label{eq:profile}
    M(>\gamma\beta)= M_0
    \begin{cases}
    \left(\frac{\gamma\beta}{\gamma_0\beta_0}\right)^{-s_\text{ft}} & \text{for}\quad \gamma_0\beta_0<\gamma\beta,\\
    \left(\frac{\gamma\beta}{\gamma_0\beta_0}\right)^{-s_\text{KN}} & \text{for}\quad 0.1<\gamma\beta<\gamma_0\beta_0,
    \end{cases}
\end{equation}
with parameter values characteristic of the results of numerical calculations of the BNS ejecta; $0.3<\beta_0<0.5$, $5<s_\text{ft}<12$, $0.5<s_\text{KN}<3$, and $10^{-6}<M_0<10^{-4}$. This analytic form provides a good approximation for the variety of ejecta profiles obtained in NR simulations of BNS mergers \citep{zappa_binary_2023}. The effects of deviations from spherical symmetry, which may be significant for mergers of objects with a large, $>1.5$, mass ratio, will be discussed in a follow-up paper under preparation.

We show in appendix~\ref{app:E} that for the case of a broken power-law dependence of ejecta energy on momentum,
\begin{equation}
\label{eq:profileE}
    E(>\gamma\beta)= E_0
    \begin{cases}
    \left(\frac{\gamma\beta}{\gamma_0\beta_0}\right)^{-\alpha_\text{ft}} & \text{for}\quad \gamma_0\beta_0<\gamma\beta,\\
    \left(\frac{\gamma\beta}{\gamma_0\beta_0}\right)^{-\alpha_\text{KN}} & \text{for}\quad 0.1<\gamma\beta<\gamma_0\beta_0,
    \end{cases}
\end{equation}
accurate analytic approximations are obtained by substituting (similar to \citetalias{sadeh_non-thermal_2023})
\begin{equation}
\begin{aligned}
\label{eq:Eofv}
  M_0&=1.5\frac{E_0}{(\gamma_0\beta_0)^{2}c^2},\\
  s_{\rm ft}&=\alpha_{\rm ft}+2.
  \end{aligned}
\end{equation}

The non-thermal flux is derived assuming that fractions $\varepsilon_e$ and $\varepsilon_B$ of the post-shock internal energy density are carried by non-thermal electrons and magnetic fields, respectively, and assuming that the electrons are accelerated to a power-law energy distribution\footnote{We do not consider the effects of a possible thermal electron component, which may be significant for large ISM densities ($\sim0.1$cm$^{-3}$) or at early times ($t<100$d) \citep{nedora_modelling_2023}.}, $dn_e/d\gamma_e\propto \gamma_e^{-p}$, where $\gamma_e$ is the electron Lorenz factor (at the plasma rest frame) and $2\le p \le2.5$. This phenomenological description of the post-shock plasma conditions is supported by a wide range of observations and plasma calculations, for both relativistic and non-relativistic shocks \citep[see][]{blandford_particle_1987,waxman_gamma-ray_2006,bykov_fundamentals_2011,sironi_maximum_2013,pohl_pic_2020,ligorini_mildly_2021,kobzar_electron_2021}. 

This paper is organized as follows. In \S~\ref{sec:analytic}, we provide the analytic derivation of the observed emission ring radius, of the self-absorption frequency, and of the cooling frequency. In \S~\ref{sec:numerical} (and appendix~\ref{app:numeric}), we describe the numerical calculation scheme. The accuracy of the analytic formulae is determined by comparisons to the results of numeric calculations in \S~\ref{sec:validation}. A comparison to earlier results is given in \S~\ref{sec:nakar}. A concise summary is given in \S~\ref{sec:conclusions}.

\section{Analytic approximations}
\label{sec:analytic}
\subsection{Image radius}
\label{sec:image_radius}
As the ejected plasma propagates into the ISM, a forward-reverse shock structure is formed; the forward shock propagates into the cold medium ahead, while the reverse shock propagates into the ejecta and decelerates it. Radiation emitted from an angle $\theta$ with respect to the line of sight and observed at time $t$ was emitted when the forward shock radius was (\citetalias{sadeh_non-thermal_2023})
\begin{equation}
\label{eq:R}
    R(t,\theta)=\frac{\beta_{\rm ej,RS} ct}{1-\beta_{\rm ej,RS}\cos\theta},
\end{equation}
where $\beta_{\rm ej,RS} (\gamma_{\rm ej,RS})$ is the initial velocity (Lorentz factor) of the ejecta shell
that was reached by the reverse shock (at the time the forward shock reached $R$).
The perpendicular distance of the emitting plasma from the line of sight is given by $R_\perp(\theta)\equiv R(t,\theta)\sin\theta$. For a spherical explosion, the outer edge of the observed image is a circle, and its radius is obtained by solving $\partial_\theta R_\perp(\theta)=0$. The solution is $\theta=\theta_\gamma$, with
\begin{equation}
\label{eq:theta_g}
    \cos\theta_\gamma=\beta_{\rm ej,RS},\quad \sin\theta_\gamma=\gamma_{\rm ej,RS}^{-1},
\end{equation}
and 
\begin{equation}
   R_\perp(\theta_\gamma)= \gamma_{\rm ej,RS}\beta_{\rm ej,RS} ct.
\end{equation}

In \citetalias{sadeh_non-thermal_2023} we showed that 
\begin{equation}
\label{eq:gam}
    \gamma_{\rm ej,RS}\beta_{\rm ej,RS}=\left(\frac{t}{t_R}\right)^{-\frac{3}{5.5+s_\text{ft}}},
\end{equation}
where 
\begin{equation}
\label{eq:mRtR}
    t_R\equiv \left(\frac{M_R}{16 \pi nm_pc^3}\right)^\frac{1}{3} ,\quad
     M_R\equiv M(\gamma\beta>1)= M_0(\gamma_0\beta_0)^{s_{\rm ft}}.
\end{equation}
$M_R$ is the "relativistic mass" with momentum $\gamma\beta>1$, and $t_R$ is approximately the time at which the reverse shock crosses $M_R$. Using these results, we have
\begin{equation}
\label{eq:ring_radius}
R_\perp(\theta_\gamma)=1.05ct_R\left(\frac{t}{t_R}\right)^{\frac{2.5+s_\text{ft}}{5.5+s_\text{ft}}},
\end{equation}
where the $1.05$ factor is obtained by fitting to the results of numerical calculations.

At the peak time (see Eq. (23) in \citetalias{sadeh_non-thermal_2023}),
\begin{equation}
   \label{eq:tpeak}
   t_\text{peak}= 550g(\beta_0)\left(\frac{M_{0,-4}}{n_{-2}}\right)^{\frac{1}{3}}\text{days},\quad
   g(\beta_0)=\frac{1.5-\sqrt{0.25+2\beta_0^2}}{\gamma_0^{\frac{1}{3}}\beta_{0}}, 
\end{equation}
and for typical parameters ($\gamma_0\beta_0\approx0.3,s_\text{ft}=7$), the image angular scale is given by  
\begin{equation}
\label{eq:eta}
    \eta =\frac{2R_\perp(\theta_\gamma)}{D} = 6.4\left(\frac{M_{R,-6}}{n_{-2}}\right)^{\frac{1}{3}}D_{26.5}^{-1} \text{mas},
\end{equation}
where $M_R\equiv10^{-6} M_{R,-6}M_\odot$, $n=10^{-2}n_{-2}{\rm cm}^{-3}$ and $D=10^{26.5}D_{26.5}{\rm cm}$.

\subsection{Self-absorption frequency}
\label{sec:self-absorption}

We estimate the self-absorption frequency at time $t$ as the frequency for which $\tau_\nu=\alpha_\nu\Delta_\tau=1$, where $\alpha_\nu$ and $\Delta_\tau$ are the typical absorption coefficient and the typical path length traversed by photons through the shocked plasma, when the forward shock reached the radius $R$, dominating the emission of radiation observed at time $t$. 

The contribution to the observed flux from emission at shock radius $R$ is dominated by plasma located at an angle 
\begin{equation}
\label{eq:cos}
    \cos\theta= -\frac{1}{2\beta}+\sqrt{2+\frac{1}{4\beta^2}},
\end{equation} where $\beta$ is the velocity of the shocked plasma right behind the forward shock at radius $R$ (\citetalias{sadeh_non-thermal_2023}). Noting that the post-shock density is approximately $4\gamma^2$ times the pre-shock density, we estimate the thickness $\Delta_R$ of the emitting layer as
    \begin{equation}
\begin{aligned}
    \frac{4\pi}{3}R^3&\approx\frac{4\pi}{3}4\gamma^2\left(R^3-(R-\Delta_R)^3\right),  \\
    \Delta_R&\approx R\left(1-\frac{\left(4\gamma^2-1\right)^\frac{1}{3}}{\left(4\gamma^2\right)^\frac{1}{3}}\right).
\end{aligned}
\end{equation}
Using Eqs.~\ref{eq:cos} and  \ref{eq:R}, and approximating $\beta\approx \beta_{\rm ej,RS}$, we have
\begin{equation}
    R(t)=\frac{\beta ct}{\left(1.5-\sqrt{2\beta^2+0.25}\right)},
\end{equation}
and
\begin{equation}
\begin{aligned}
\Delta_\tau&\approx\frac{\Delta_R}{\cos\theta}\approx\frac{R}{\cos\theta}\left(1-\frac{\left(4\gamma^2-1\right)^\frac{1}{3}}{\left(4\gamma^2\right)^\frac{1}{3}}\right)\\&\approx\frac{ \beta^2 ct}{\left(1.5-\sqrt{2\beta^2+0.25}\right)\left(-0.5+\sqrt{2\beta^2+0.25}\right)}\left(1-\frac{\left(4\gamma^2-1\right)^\frac{1}{3}}{\left(4\gamma^2\right)^\frac{1}{3}}\right).
\end{aligned}
\end{equation}

To derive the self-absorption frequency, we first approximate the absorption coefficient Using the results from \citetalias{sadeh_non-thermal_2023} for the post-shock magnetic field strength, $B=\sqrt{32\pi\varepsilon_B \gamma(\gamma-1)nm_pc^2}$, the electron energy distribution, $dn_e/d\gamma_e= (p-1)\left(\frac{p-2}{p-1}\frac{\frac{\varepsilon_e m_p}{m_e}(\gamma-1) }{1-\left(\gamma_\text{max}/\gamma_\text{min}\right)^{2-p}}\right)^{p-1}4\gamma n\gamma_e^{-p}$ ($\gamma_\text{max},\gamma_\text{min}$ are the maximal, minimal electron Lorentz factors respectively), and the relation $\nu'/\nu\approx\gamma\left(1.5-\sqrt{2\beta^2+0.25}\right)$ the frequency in the rest frame of the emitting plasma and the observed frequency, the absorption coefficient is given by \citep[using the results of][for power-law electron distribution]{rybicki_radiative_1979}
\begin{equation}
    \alpha_\nu = \frac{\nu'}{\nu}\alpha_\nu'\approx \gamma^{\frac{2-p}{4}}\left(1.5-\sqrt{2\beta^2+0.25}\right)^{-\frac{p+2}{2}}(\gamma-1)^{\frac{5p-2}{4}}f_a(p)\nu^{-\frac{p+4}{2}},
\end{equation}
where 
\begin{equation}
\begin{aligned}
       &f_a(p)=(p-1)\left(\frac{p-2}{p-1}\frac{\frac{\varepsilon_e m_p}{m_e} }{1-\left(\gamma_\text{max}/\gamma_\text{min}\right)^{2-p}}\right)^{p-1}4n \left(\frac{2\pi m_e c}{3q_e}\right)^{-\frac{p}{2}} \times\\
       &\frac{(32\pi\varepsilon_B nm_p c^2)^{\frac{p+2}{4}}\sqrt{3}q_e^3}{8\pi m_e^2c^2}\frac{\sqrt{\pi}}{2}\frac{\Gamma\left(\frac{p+6}{4}\right)}{\Gamma\left(\frac{p+8}{4}\right)}\Gamma\left(\frac{3p+22}{12}\right)\Gamma\left(\frac{3p+2}{12}\right). 
\end{aligned}
\end{equation}

Following the same power-law approximation used in \citetalias{sadeh_non-thermal_2023}, appendix A, the optical depth at (observed) frequency $\nu$ is given by
\begin{equation}
    \tau_\nu\approx \alpha_\nu\Delta_\tau\approx(0.3-0.1p)(\gamma\beta)^{2.5p+1}f_a(p)ct\nu^{-\frac{p+4}{2}},
\end{equation}
and can be approximated as (by Eq. (\ref{eq:gam}))
\begin{equation}
    \tau_\nu\approx (0.3-0.1p)ct_Rf_a(p)\left(\frac{t}{t_R}\right)^{\frac{s_\text{ft}+2.4-7.7p}{5.5+s_\text{ft}}}\nu^{-\frac{p+4}{2}}.
\end{equation}
The self-absorption frequency, $\nu_{a}$ defined by $\tau_\nu(\nu=\nu_{a})=1$, is finally given by
\begin{equation}
    \nu_{a,\rm ft}= 1.2\left((0.3-0.1p)ct_Rf_a(p)\left(\frac{t}{t_R}\right)^{\frac{s_\text{ft}+2.4-7.7p}{5.5+s_\text{ft}}-\frac{3}{5.5+s_\text{ft}}}\right)^{\frac{2}{p+4}},
\end{equation}
where the ft subscript implies (following the notation of \citetalias{sadeh_non-thermal_2023}) that the result is valid as long as the reverse shock propagates through the fast tail. The $1.2$ pre-factor and the $-\frac{3}{5.5+s_\text{ft}}$ power-law index are obtained from fitting to the results of numeric calculations.

For $p=2.2$ we find
\begin{equation}
\label{eq:nu_a}
    \nu_{a,\rm ft}=33\varepsilon_{e,-1}^\frac{2(p-1)}{p+4}\varepsilon_{B,-2}^{\frac{p+2}{2(p+4)}}n_{-2}^{\frac{3p+14}{6(p+4)}}M_{R,-6}^{\frac{2}{3(p+4)}}\left(\frac{t}{t_R}\right)^\frac{2s_{\rm ft}-1.2-15.4p}{(5.5+s_{\rm ft})(p+4)}\text{MHz},
\end{equation}
where $\varepsilon_{e}=10^{-1}\varepsilon_{e,-1}$ and $\varepsilon_{B}=10^{-2}\varepsilon_{B,-2}$. The numeric value of the pre-factor varies by $\pm20\%$ for $p$ values in the range $2<p<2.5$.

\subsection{Cooling frequency}
\label{sec:cooling}
In \citetalias{sadeh_non-thermal_2023}, we estimated the cooling frequency as a function of observed time. 
It is defined as the (observer frame) synchrotron frequency of electrons for which the (plasma frame) synchrotron loss time, $m_ec^2/(\gamma_e(4/3) \sigma_T cu_B)$, is equal to the time measured at the plasma frame, $t/\gamma/\left(1.5-\sqrt{2\beta^2+0.25}\right)$.
Here, we numerically calibrate the values of the pre-factor and temporal power-law index of the result given in \citetalias{sadeh_non-thermal_2023} to better fit the results of numerical calculations, obtaining
\begin{equation}
\begin{aligned}
\label{eq:nu_c}
         \nu_{c,\rm ft}&= 2.7\times 10^{9}\varepsilon_{B,-2}^{-\frac{3}{2}}n_{-2}^{-\frac{5}{6}}M_{R,-6}^{-\frac{2}{3}}\left(\frac{t}{t_R}\right)^{\frac{2.5-2s_{\rm ft}}{5.5+s_{\rm ft}}}\text{GHz}.
\end{aligned}
\end{equation}
The dependence on $\{\varepsilon_B,n,M_R\}$ is the same as that obtained in \citetalias{sadeh_non-thermal_2023}.

\section{Numerical Calculations}
\label{sec:numerical}
\subsection{Hydrodynamics}
For an accurate description of the dynamics, we employ the hydrodynamic code described in \citetalias{sadeh_non-thermal_2023}, which solves the full special relativistic 1D hydrodynamics equations. Our numerical 1D Lagrangian code utilizes a predictor-corrector scheme, incorporates artificial viscosity, and sets time steps using Courant–Friedrichs–Lewy (CFL) conditions. The code was validated by rigorous comparisons of its results to well-established solutions of benchmark problems \citep{sedov_propagation_1946, marti_numerical_2003, guzman_revisiting_2012}.
The equation of state we use is $p=(\hat{\gamma}-1)e$, where the adiabatic index $\hat{\gamma}(e/n)$ varies within the range of $4/3$ to $5/3$, following  \cite{synge_relativistic_1957}.
The initial ejecta density and velocity profiles are given in Eq. (\ref{eq:profile}), and the initial, $t=t_0$, radius of a mass shell $M(>\gamma\beta)$ is determined by $r_0=\beta ct_0$. The expanding ejecta is initially (at $t=t_0$) embedded (at $r>c t_0$) in a static, uniform, cold (zero pressure) gas with number density $n$. The ejecta comprises 700 numeric cells out of a total of 5000 cells. Convergence is verified by doubling the resolution in the radial direction and reducing the time steps by a factor of 2, yielding indistinguishable results. 

\subsection{Radiation}
\label{sec:numerical_flux}
To calculate the emission of radiation, noting the azimuthal symmetry, we use cylindrical coordinates, $r$ and $z$, where the $z$ axis is aligned with the line of sight, such that the radiation emitted by a cell at $t_\text{lab}$ arrives to the observer at
\begin{equation}
    t_\text{obs}=t_\text{lab}-\frac{z}{c}.
\end{equation}
We divide space into $N_r\times N_z$ ($N_r,N_z\sim$ few $1000$ each) cells. Using the flow fields ($\rho,\gamma,e$) obtained in the hydrodynamic calculations, the synchrotron emission is calculated in the following manner (full details are given in appendix \ref{app:numeric}): 
We assume that fractions $\varepsilon_e$ and $\varepsilon_B$ of the internal energy density are held by magnetic fields and electrons and that the electrons are accelerated to a power-law distribution $dn_e/d\gamma_e\propto \gamma_e^{-p}$.
The emissivity, $j_\nu$, and the absorption coefficient, $\alpha_\nu$, in each location and time, are defined in the plasma frame, and then Lorentz boosted to the lab frame. 

To consider the effect of synchrotron cooling of high energy electrons, we numerically solve (as described in appendix~\ref{app:numeric}) the energy evolution of the electrons, determine the (spatially and temporally) dependent Lorentz factor $\gamma_c$ above which the electron distribution is strongly suppressed by cooling, and calculate the radiation emitted by a power-law energy distribution of electrons (properly shifted down in energy due to adiabatic expansion) truncated at $\gamma_c$. 

The contribution to the observed intensity of each cell is given by
\begin{equation}
\label{eq:numeric_nua}
    \Delta I_\nu(t_\text{obs}) = \frac{j_\nu\left(t_\text{obs}+\frac{z}{c}\right)}{\alpha_\nu\left(t_\text{obs}+\frac{z}{c}\right)}\left(\Delta \tau\left(t_\text{obs}+\frac{z}{c}\right)\right)e^{-\tau_\nu\left(t_\text{obs}+\frac{z}{c}\right)},
\end{equation}
where $\Delta \tau_\nu=\alpha_\nu \Delta z$ and $\tau_\nu$ is the optical depth along the path to the observer
(In calculating $\tau_\nu\left(t_\text{obs}+\frac{z}{c}\right)$ we take into account the evolution of the shocked plasma during the photons' propagation through it, with properly evaluated $\alpha_\nu\left(t=t_\text{obs}+\frac{z}{c}\right)$ along the path). To accelerate numeric convergence, the factor $\left(\Delta \tau\left(t_\text{obs}+\frac{z}{c}\right)\right)$ in Eq. (\ref{eq:numeric_nua}) is replaced with  $\left(1-e^{-\Delta \tau_\nu\left(t_\text{obs}+\frac{z}{c}\right)}\right)$. 

The flux and fraction of the flux originating from different annuli are defined as
\begin{equation}
\label{eq:flux}
\begin{aligned}
    F_\nu&=\frac{2\pi}{D^2}\int r I_\nu dr,\\ 
    \frac{df_\nu}{dr}&=\frac{\frac{2\pi}{D^2}r I_\nu}{F_\nu},
    \end{aligned}
\end{equation}
where $D$ is the distance to the observer.

The convergence of our numeric radiation calculations is verified by doubling the resolution in both $r$ and $z$ axes, yielding indistinguishable results.

\section{Comparison of analytic and numeric results}
\label{sec:validation}
In comparing our numeric and analytic results, we have explored a wide range of values of the dimensionless parameters determining the hydrodynamic behavior, $5<s_{\rm ft}<10,~0.3<\beta_0<0.5$, and the range $2<p<2.4$ of the electron spectral index. The dependence of the dynamics on the dimensional parameters, $\{M_0, n, c\}$, follows directly from dimensional considerations, while the dependence on $\varepsilon_e$ and $\varepsilon_B$ is analytically obtained straightforwardly. Since we provide results for times earlier than the peak time, the time at which the reverse shock reaches the shell with initial momentum $\gamma_0\beta_0$, our results are independent of $s_{\rm KN}$.

\subsection{Sky image}
\begin{figure*}
    \centering
    \begin{subfigure}[b]{0.45\textwidth}
\includegraphics[width=\columnwidth]{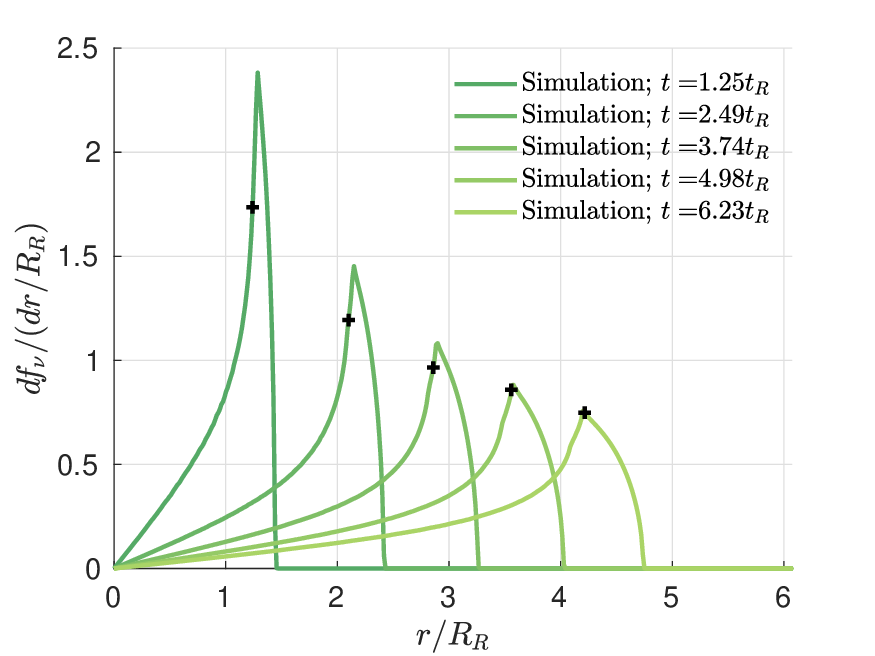}
    \end{subfigure}
    \hfill
\begin{subfigure}[b]{0.45\textwidth}
\includegraphics[width=\columnwidth]{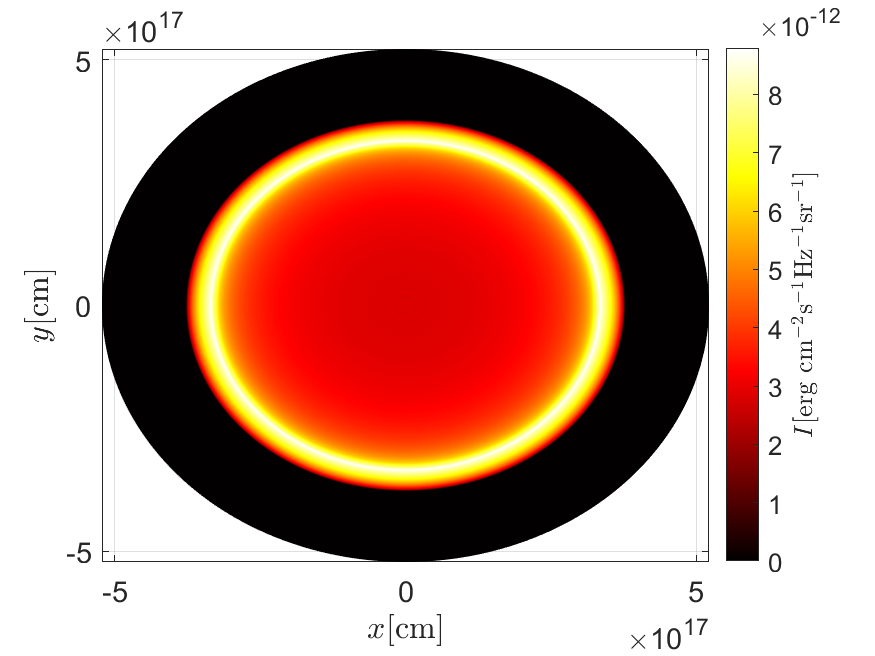}
\end{subfigure}
    \caption{Left panel: The fraction of flux that is emitted at different times by different annuli, $df_\nu/dr$, (obtained numerically) for $p=2.2,s_\text{ft}=7,\gamma_0=1.15$. $r$, the radius of the annulus (its transverse distance to the line of sight) is normalized to $R_R=ct_R$ (approximately the radius at which the post-shock plasma momentum drops to $\gamma\beta=1$), where $t_R$ is defined in Eq. (\ref{eq:mRtR}). The $+$'s are the analytic estimate of the image radius, given by Eq. (\ref{eq:ring_radius}). Right panel: A corresponding intensity map at observed time $t=300$days ($I_\nu$ and $df_\nu/dr$ are related through Eq. \ref{eq:flux}).
    }
    \label{fig:ang}
\end{figure*}

\begin{figure}
    \centering
\includegraphics[width=\columnwidth]{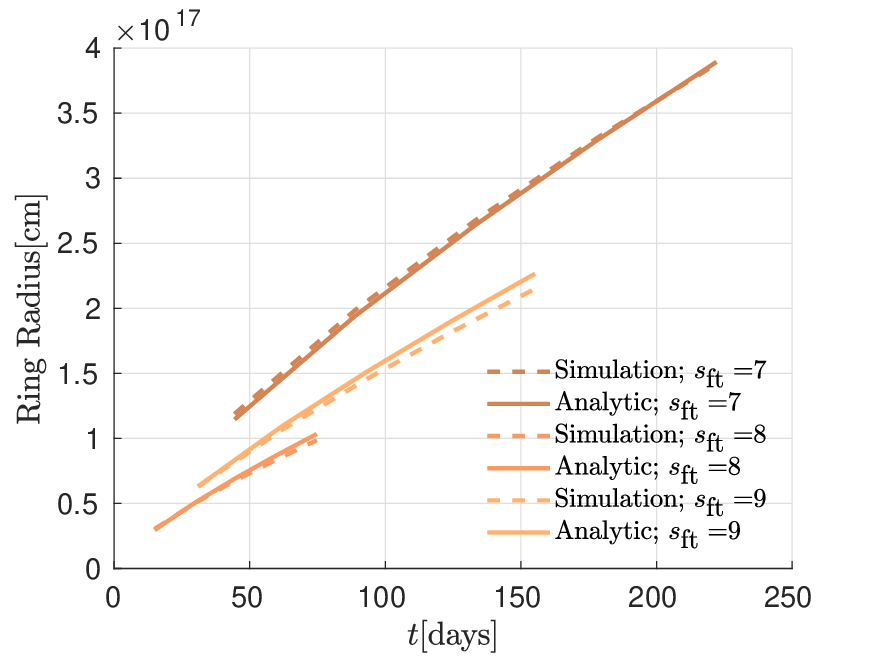}
    \caption{A comparison between the analytic (solid lines, Eq. (\ref{eq:ring_radius})) and numeric (dashed line) time-dependent image radii for different sets of values of the parameters $\{s_\text{ft},\gamma_0, M_{R,-6},n_{-2}\}$: $\{7,1.15,1,3\},\,\{8,1.09,10^{-3},3\},\,\{9,1.1,10^{-2},3\}$.}
     \label{fig:image_radius}
\end{figure}

In Fig.~\ref{fig:ang}, we show an example of the flux emitted from different annuli obtained numerically along with the analytic estimate of the image radius, Eq.~ (\ref{eq:ring_radius}). Due to relativistic beaming and time arrival effects, the image is a relatively narrow ring.
In Fig.~\ref{fig:image_radius}, we compare the image radius obtained numerically (defined by the radial position of the peak of $df_\nu/dr$) with the analytic estimate, Eq.~(\ref{eq:ring_radius}). The agreement is to within a few percent for a wide range of relevant values of $\{\beta_0,s_{\rm ft}\}$.

\subsection{Spectrum}
\begin{figure}
    \centering
\includegraphics[width=\columnwidth]{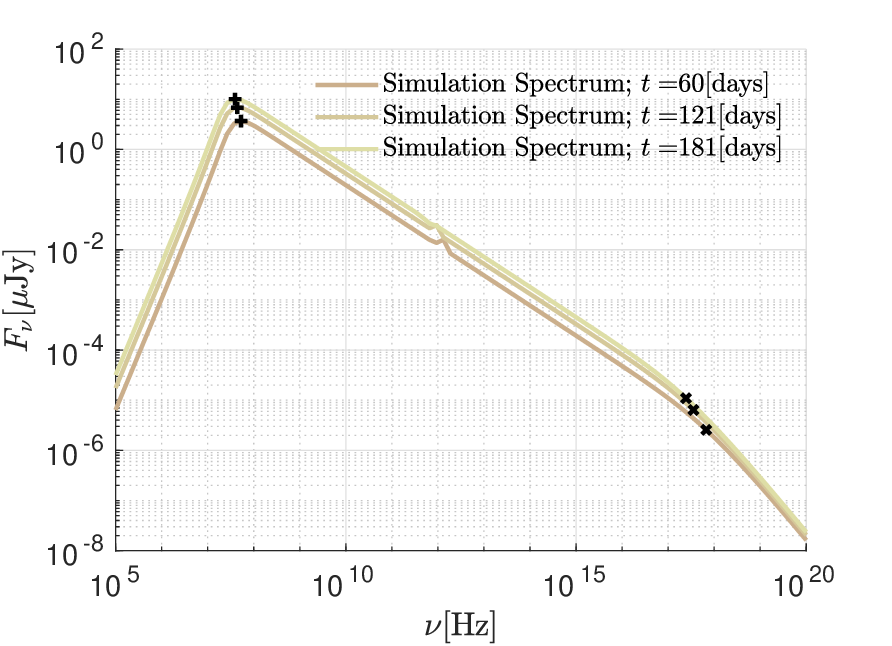}   
    \caption{The numerically calculated spectra at three epochs for $p=2.2,s_\text{ft}=7,\varepsilon_e=10^{-1},\varepsilon_B=10^{-2},M_R=10^{-6}M_\odot,n=3\times10^{-2}\text{cm}^{-3}$, normalized for a distance of $100$Mpc. The analytic self-absorption and cooling frequencies, Eqs. (\ref{eq:nu_a}-\ref{eq:nu_c}), are shown by '+' and 'x' signs respectively.}
    \label{fig:spectrum}
\end{figure}

\begin{figure}
    \centering
\includegraphics[width=\columnwidth]{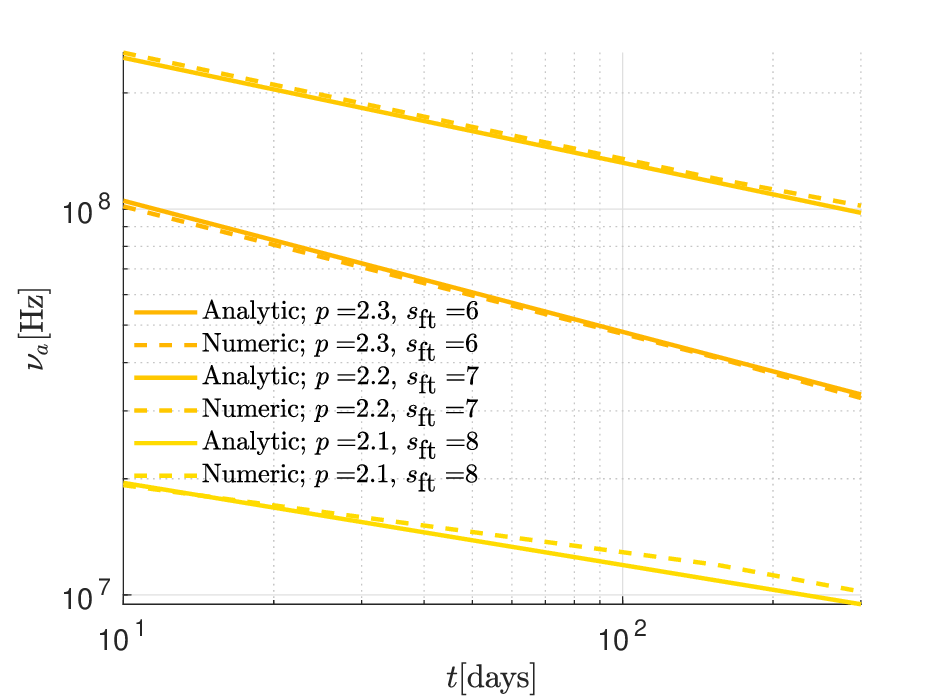}
    \caption{A comparison of the analytic (Eq.~(\ref{eq:nu_a}), solid lines) and numeric (dashed lines) self-absorption frequency, $\nu_a$, for different sets of values of $\{s_\text{ft},p, M_{R,-6},n_{-2},\varepsilon_{B,-2},\varepsilon_{e,-1}\}$: $\{6,2.3,1,3,1,1\},\,\{7,2.2,27.3,7,1,1\},\,\{8,2.1,10^{-3},3,1,1\}$.}
    \label{fig:self-absorption}
\end{figure}

\begin{figure}
    \centering
\includegraphics[width=\columnwidth]{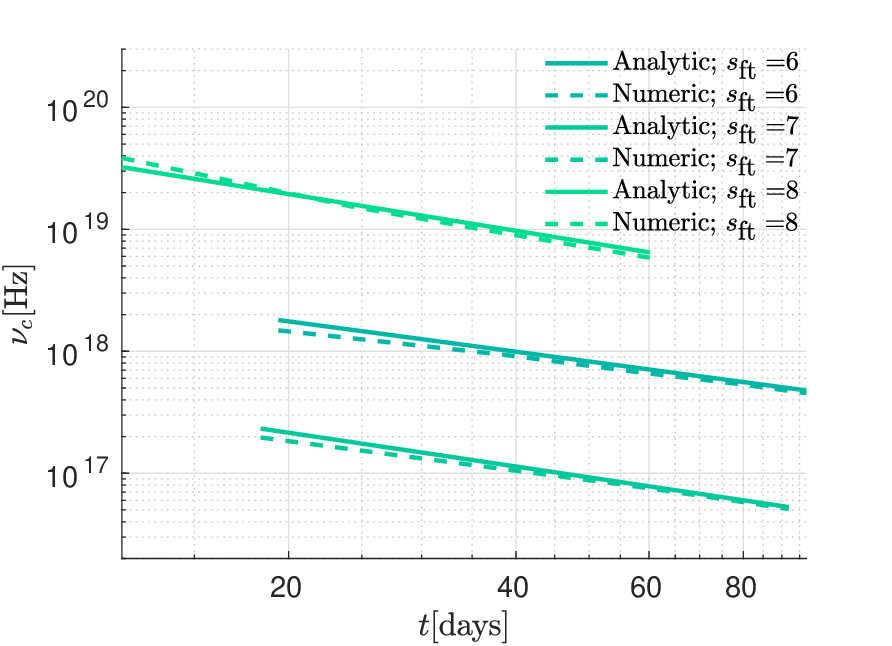}
    \caption{A comparison of the analytic (Eq.~(\ref{eq:nu_c}), solid lines) and numeric (dashed lines) self-absorption frequency, $\nu_c$, for different sets of values of $\{s_\text{ft},M_{R,-6},n_{-2},\varepsilon_{B,-2}\}$: $\{6,1,3,1\},\,\{7,27.3,7,1\},\,\{8,10^{-3},3,1\}$.}
    \label{fig:cooling}
\end{figure}

Fig. {\ref{fig:spectrum}} presents an example of spectra obtained by our numeric calculations, compared with the analytic estimates of the self-absorption and cooling frequencies, Eqs.~(\ref{eq:nu_a}) and~(\ref{eq:nu_c}) respectively. In Figs.~\ref{fig:self-absorption} and~\ref{fig:cooling}, we compare the cooling and self-absorption frequencies obtained numerically with those given by the analytic approximations, Eqs.~(\ref{eq:nu_a}) and~(\ref{eq:nu_c}). We define the self-absorption[cooling] frequency in the numerical calculation as the minimal frequency at which the spectrum frequency dependence reaches $d\log(F_\nu)/d\log(\nu)\leq0[(1-2p)/4]$. The values $0$ and $(1-2p)/4$ are chosen as they are intermediate values between those of the different spectral regimes: 
\begin{equation}
    \begin{aligned}
        F_\nu\propto \begin{cases}
           \nu^{5/2},\quad \text{for } \nu<\nu_a,\\
           \nu^{(1-p)/2},\quad \text{for } \nu_a<\nu<\nu_c,\\
           \nu^{-p/2},\quad \text{for } \nu_c<\nu.
        \end{cases}
    \end{aligned}
\end{equation}
The agreement is within $10$'s of percent for a wide range of relevant values of $\{\beta_0,s_{\rm ft}\}$.

\section{Comparison to an earlier work}
\label{sec:nakar}
\cite{nakar_radio_2011} derived an order of magnitude estimate of the self-absorption frequency at the peak time, extrapolating results valid for $\beta\ll1$, 
\begin{equation}
\nu^{\text{NP}}_{a,\text{peak}}=1\varepsilon_{e,-1}^\frac{2(p-1)}{p+4}\varepsilon_{B,-1}^{\frac{p+2}{2(p+4)}}n_{-2}^{\frac{3p+14}{6(p+4)}}E_{49}^{\frac{2}{3(p+4)}}\beta_0^\frac{15p-10}{2(p+4)}\text{GHz},
\end{equation}
where the energy of the fast tail is $E_0=10^{49}E_{49}$erg, and the numeric coefficient (1) was obtained for $p=2.5$. Using our result, Eq. (\ref{eq:nu_a}), at $t=t_\text{peak}$ (given by Eq.~(\ref{eq:tpeak})), and the relation of Eq.~(\ref{eq:Eofv}) between $E_0$ and $M_0$, we find 
\begin{equation}
\nu_{a,\text{peak}}=0.1\varepsilon_{e,-1}^\frac{2(p-1)}{p+4}\varepsilon_{B,-1}^{\frac{p+2}{2(p+4)}}n_{-2}^{\frac{3p+14}{6(p+4)}}E_{49}^{\frac{2}{3(p+4)}}\left(\gamma_0\beta_0\right)^{\frac{-1.9+10.2p}{2(p+4)}}\text{GHz}
\end{equation}
(the numeric coefficient (0.1) is obtained for $\alpha_{\rm ft}=5$ and $p=2.5$). The ratio between our (numerically verified) analytic result and that of \cite{nakar_radio_2011} is
\begin{equation}
    \frac{\nu^{\text{NP}}_{a,\text{peak}}}{\nu_{a,\text{peak}}}= 10\gamma_0 ^{\frac{1.9-10.2p}{2(p+4)}}\beta_0^\frac{4.8p-8.1}{2(p+4)}.
\end{equation}
For typical values of $\beta_0=0.3$, the order of magnitude estimate of \cite{nakar_radio_2011} is accurate to within a factor of $\sim5$.

\section{Conclusions}
\label{sec:conclusions}
Analytic expressions were derived in \S~\ref{sec:analytic} for the image radius, the self-absorption frequency, $\nu_a$, and the cooling frequency, $\nu_c$, of the non-thermal emission from a collisionless shock driven into the ISM by mildly relativistic spherical ejecta with broken power-law mass or energy distributions, given by Eqs.~(\ref{eq:profile}) or~(\ref{eq:profileE}), at times earlier than the peak time of the emission at $\nu_a<\nu<\nu_c$ (Eqs. (\ref{eq:ring_radius}), (\ref{eq:nu_a}) and (\ref{eq:nu_c})). The analytic model results were compared in \S~\ref{sec:validation} to the results of 1D numeric calculations for a wide range of ejecta parameter values characteristic of merger calculation results, $5<s_{\rm ft}<10,~0.3<\beta_0<0.5$, and for the range $2<p<2.4$ of the electron spectral index (The dependence of the dynamics on the dimensional parameters, $\{M_0, n, c\}$, follows directly from dimensional considerations, while the dependence on $\varepsilon_e$ and $\varepsilon_B$ is analytically obtained straightforwardly). We showed that the analytic model expressions reproduce the results of numeric calculations with tens of percent accuracy; see Figs.~\ref{fig:ang}-\ref{fig:cooling}. This is a significant improvement over earlier order-of-magnitude estimates, based on extrapolations of results valid for $\gamma\beta\ll1$ to $\gamma\beta\approx1$. 

The results presented in this paper, combined with the analytic results derived in \citetalias{sadeh_non-thermal_2023} for the peak time, peak flux, and temporal power-law indices of the flux rise and decline, 
will enable to constrain the parameters of the model, $\{s_\text{ft}, s_\text{KN}, M_0, n, \gamma_0\beta_0, \varepsilon_e, \varepsilon_B, p\}$, using future observational data. We note that an analytic description is essential for model parameter inference from data since the model depends non-trivially on several parameters, $\{s_\text{ft}, s_\text{KN}, \gamma_0\beta_0, p\}$, and numerical calculations of the model predictions for each set of parameter values requires significant computational resources and time.

The spectral flux $F_\nu$ peaks at $\nu\sim\nu_a$, typically in the range of $10-100$~MHz, and suppressed beyond $\nu_c$, typically at $\sim 10^{10}$~GHz. For sources at a distance of $\sim100$~Mpc, the VLA may measure the self-absorption frequency, while the cooling frequency may be determined using Chandra X-ray data. 

In \citetalias{sadeh_non-thermal_2023}, we have shown that the existence of a "fast dynamical tail" in the ejecta associated with GW170817 is expected to produce detectable (few $\mu$Jy) radio and
($10^{-15}\rm{erg\, cm^{-2} /s}$) X-ray fluxes over a time scale of $\sim 10^4$~d. The image is predicted to reach on this time scale an angular size exceeding 10~mas (Eqs.~(\ref{eq:ring_radius}), (\ref{eq:eta})), which is resolvable by VLBI. 

\section*{Acknowledgements}
Eli Waxman's research is partially supported by ISF and GIF grants.

\section*{Data Availability}
The data underlying this article will be shared following a reasonable request to the corresponding author.


\bibliographystyle{mnras}
\bibliography{references} 




\appendix
\section{broken power-law dependence of ejecta energy on momentum}
\label{app:E}
\begin{figure}
    \centering
\includegraphics[width=\columnwidth]{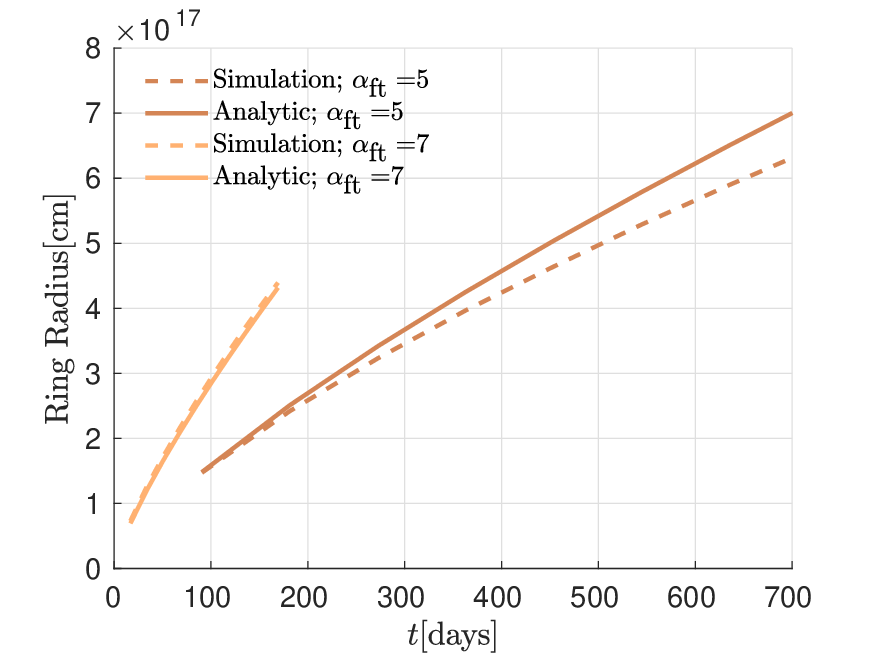}
    \caption{A comparison between the analytic (solid lines, Eq. (\ref{eq:ring_radius}) using the substitutions of Eq. \ref{eq:Eofv}), and numeric (dashed lines) time-dependent image radii for different sets of values of the parameters $\{\alpha_\text{ft},\gamma_0,E_{0}/10^{49}\text{erg},n_{-2}\}$: $\{5,1.05,1,3\},\,\{7,1.35,10,3\}$.}
     \label{fig:image_radius_E}
\end{figure}

\begin{figure}
    \centering
\includegraphics[width=\columnwidth]{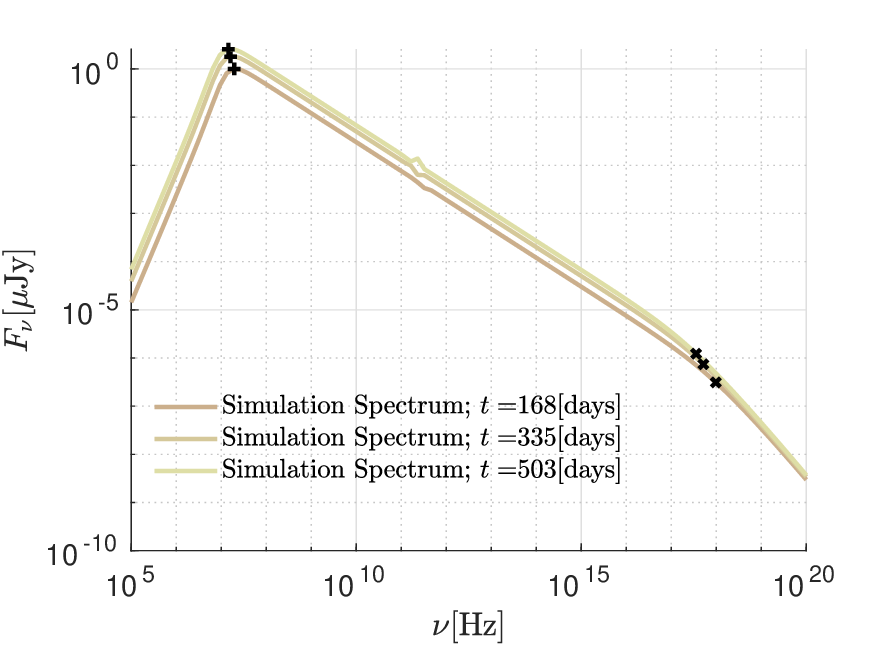}   
    \caption{The numerically calculated spectra at three epochs for $p=2.2,\alpha_\text{ft}=5,\varepsilon_e=10^{-1},\varepsilon_B=10^{-2},E_0=10^{49}\text{erg},n=3\times10^{-2}\text{cm}^{-3}$, for a source distance of $100$Mpc. The analytic self-absorption and cooling frequencies, Eqs. \ref{eq:nu_a} and \ref{eq:nu_c} using the substitutions of Eq. \ref{eq:Eofv}, are shown by '+' and 'x' symbols.}
    \label{fig:spectrum_E}
\end{figure}

Analytic expressions for the synchrotron emission produced by ejecta with a power-law dependence of ejecta energy on momentum, Eq.~(\ref{eq:profileE}), may be obtained from the results given in \S \ref{sec:analytic} by the substitutions of Eq.~(\ref{eq:Eofv}). The validity of the resulting analytic expressions is demonstrated in Figs. \ref{fig:image_radius_E} and \ref{fig:spectrum_E}, comparing the analytic results with numeric results for an ejecta described by Eq.~(\ref{eq:profileE}).

\section{Numerical scheme for synchrotron emission calculation}
\label{app:numeric}
We begin by defining the emissivity and absorption from the flow fields ($\rho(R,t),\gamma(R,t),e(R,t)$) at radius $R$ and lab time $t$ (time in the rest frame of the ISM). The synchrotron power of a single electron is \citep{rybicki_radiative_1979}
\begin{equation}
    P_\nu = \frac{\sqrt{3}q_e^3 B\sin\alpha}{m_ec^2}F(x),
\end{equation}
where $\alpha$ is the angle between the magnetic field and the electron velocity, $x\equiv \frac{\nu}{\nu_c}$, $\nu_c\equiv \frac{3\gamma_e^2q_eB\sin\alpha}{4\pi m_e c}$, and $F(x)\equiv x\int_x^\infty K_{5/3}(\xi)d\xi$.
The emissivity and the absorption coefficient are given by
\begin{equation}
\begin{aligned}
    j_\nu'&=\int_{\gamma_\text{min}}^{\gamma_\text{max}} \frac{P_\nu}{4\pi}\frac{dn}{d\gamma_e}d\gamma_e,\\
    \alpha_\nu'&=-\frac{1}{8\pi m_e\nu^2}\int_{\gamma_\text{min}}^{\gamma_\text{max}} P_\nu \gamma_e^2\partial_{\gamma_e}\left(\frac{1}{\gamma_e^2}\frac{dn}{d\gamma_e}\right)d\gamma_e,
\end{aligned}
\end{equation}
where the ' denotes quantities measured in the fluid rest frame. 
Fermi acceleration is expected to produce electrons in a power-law distribution of the form $dn_e/d(\gamma_e\beta_e)=\kappa(\gamma_e\beta_e)^{-p}$, which reduces into $dn_e/d\gamma_e=\kappa\gamma_e^{-p}$ for ultra-relativistic electrons.  
For this distribution of electrons, between $\gamma_\text{min}$ and $\gamma_\text{max}$, we find
\begin{equation}
\begin{aligned}
    j_\nu'=&\frac{\sqrt{3}q_e^3}{8\pi m_ec^2}\left(\frac{4\pi m_e c}{3q_e}\right)^{\frac{1-p}{2}}\kappa B'^{\frac{p+1}{2}}\nu'^{\frac{1-p}{2}}\langle (\sin\alpha)^{\frac{p+1}{2}}\rangle\times\\&\int_{x_\text{max}}^{x_\text{min}}x^{\frac{p-3}{2}}F(x)dx,
    \end{aligned}
\end{equation}
\begin{equation}
\begin{aligned}
    \alpha_\nu'&=\frac{(p+2)\sqrt{3}q_e^3}{16\pi m_e^2c^2}\left(\frac{4\pi m_e c}{3q_e}\right)^{-\frac{p}{2}}\kappa B'^{\frac{p+2}{2}}\nu'^{-\frac{p+4}{2}}\langle (\sin\alpha)^{\frac{p+2}{2}}\rangle\times\\&\int_{x_\text{max}}^{x_\text{min}}x^{\frac{p-2}{2}}F(x)dx.
    \end{aligned}
\end{equation}
Since the emission is isotropic in the fluid rest frame, we average over the angle between the magnetic field and the electron's velocity
\begin{equation}  \langle\left(\sin\alpha\right)^{W}\rangle=\frac{1}{2}\int_0^\pi\left(\sin\alpha^*\right)^{W+1}d\alpha^*,
\end{equation}
and
\begin{equation}
\langle\left(\sin\alpha\right)^{\frac{p+1}{2}}\rangle=\frac{\sqrt{\pi}}{2}\frac{\Gamma\left(\frac{p+5}{4}\right)}{\Gamma\left(\frac{p+7}{4}\right)},\quad  \langle\left(\sin\alpha\right)^{\frac{p+2}{2}}\rangle=\frac{\sqrt{\pi}}{2}\frac{\Gamma\left(\frac{p+6}{4}\right)}{\Gamma\left(\frac{p+8}{4}\right)}.
\end{equation}

\subsection{Magnetic field and electron energy distribution}
The magnetic field in the fluid rest frame is given by 
\begin{equation}
    u_B'=\frac{B'^2}{8\pi}=\varepsilon_Be'(R,t)\rightarrow B'=\sqrt{8\pi\varepsilon_Be'(R,t)},
\end{equation}
where $u_B$ is the magnetic field energy density.
The energy density held by electrons in the fluid rest frame is given by
\begin{equation}
\begin{aligned}
        u_e'&=\int(\gamma_e-1)m_ec^2\frac{dn_e}{d\gamma_e}d\gamma_e\\&\approx\kappa m_ec^2\int \gamma_e(\gamma_e)^{-p}d\gamma_e=\varepsilon_ee'(R,t).
        \end{aligned}
\end{equation}
$\kappa$ and $\gamma_\text{min}$ are given by
\begin{equation}
\begin{aligned}
    \frac{m_p\varepsilon_e e'}{m_ec^2\rho'}&=\frac{\int^{\gamma_\text{max}}_{\gamma_\text{min}} (\gamma_e-1)\gamma_e^{-p}d\gamma_e}{\int^{\gamma_\text{max}}_{\gamma_\text{min}} \gamma_e^{-p}d\gamma_e}\rightarrow\\ \gamma_\text{min}&\approx\frac{p-2}{p-1}\frac{\frac{m_p\varepsilon_ee'}{m_ec^2\rho'} }{1-\left(\gamma_\text{max}/\gamma_\text{min}\right)^{2-p}},\\
    \kappa&=\frac{\rho'(R,t)}{m_p}\cdot\frac{1}{\int^{\gamma_\text{max}}_{\gamma_\text{min}} \gamma_e^{-p}d\gamma_e}=(p-1)\gamma_\text{min}^{p-1}\frac{\rho'(R,t)}{m_p}.
    \end{aligned}
\end{equation}
For mildly relativistic electrons, we approximate
\begin{equation}
    \gamma_\text{min}\approx \sqrt{1+\frac{p-2}{p-1}\frac{\frac{m_p\varepsilon_ee'}{m_ec^2\rho'}} {1-\left(\gamma_\text{max}/\gamma_\text{min}\right)^{2-p}}}.
\end{equation}
\subsection{Synchrotron Cooling} 
Consider electrons accelerated by the shock at (lab) time $t_s$ to a (fluid frame) Lorentz factor $\gamma_{e,i}$. The electrons lose energy by synchrotron emission,
\begin{equation}
   \frac{d\gamma_e}{dt'}m_ec^2=-P_\text{syn}=-\frac{4}{3}\sigma_Tc\gamma_e^2\varepsilon_B e(R_L,t),
\end{equation}
where $R_L$ is the Lagrangian simulation position that follows the same simulation cell to follow its cooling.
The electron Lorentz factor $\gamma_{e,f}$ reached at (lab) time $t_0$ is determined by 
\begin{equation}
\begin{aligned}
\int_{\gamma_{e,i}}^{\gamma_{e,f}}\frac{d\gamma_e}{\gamma_e^2}&=-\frac{4\sigma_T\varepsilon_B}{3m_e c}\int_{t_s}^{t_0}\frac{e(R_L,t)}{\gamma(R_L,t)} dt,\\
    \gamma_{e,f}^{-1}-\gamma_{e,i}^{-1}&=\frac{4\sigma_T\varepsilon_B}{3m_e c}\int_{t_s}^{t_0}\frac{e(R_L,t)}{\gamma(R_L,t)} dt,\\
    \gamma_{e,f}&=\frac{1}{\gamma_{e,i}^{-1}+\frac{4\sigma_T\varepsilon_B}{3m_e c}\int_{t_s}^{t_0}\frac{e(R_L,t)}{\gamma(R_L,t)} dt}.
\end{aligned}
\end{equation}
The cooling is insignificant for $\gamma_{e,i}^{-1}\gg \frac{4\sigma_T\varepsilon_B}{3m_e c}\int_{t_s}^{t_0}\frac{e(R_L,t)}{\gamma(R_L,t)} dt$ and becomes considerable for $\gamma_{e,i}^{-1}\sim\frac{4\sigma_T\varepsilon_B}{3m_e c}\int_{t_s}^{t_0}\frac{e(R_L,t)}{\gamma(R_L,t)} dt$. Thus, we define the Lorentz factor of electrons radiating at the cooling frequency as 
\begin{equation}
    \gamma_c = \frac{3m_e c}{8\sigma_T\varepsilon_B\int_{t_s}^{t_0}\frac{e(R_L,t)}{\gamma(R_L,t)} dt}.
\end{equation} 

\subsection{Lorentz Transformations}
For the emission coefficient, we have
\begin{equation}
\begin{aligned}
            j_\nu&=\frac{\nu^2}{\nu'^2}j'_\nu=\frac{1}{\gamma(R,t)^2\left(1-\beta(R,t)\mu\right)^2}j'_\nu(\nu')\\&=\frac{\gamma(R,t)^\frac{1-p}{2}\left(1-\beta(R,t)\mu\right)^\frac{1-p}{2}}{\gamma(R,t)^2\left(1-\beta(R,t)\mu\right)^2}j'_\nu(\nu),\\
            j_\nu&=\left(\gamma(R,t)\left(1-\beta(R,t)\mu\right)\right)^{\frac{-3-p}{2}}j'_\nu(\nu),
\end{aligned}
\end{equation}
where $\mu\equiv \cos\psi$, and $\psi$ is the angle between the line of sight and the plasma velocity. 

For the absorption coefficient, we have
\begin{equation}
    \begin{aligned}
        \alpha_\nu& =\frac{\nu'}{\nu}\alpha'_\nu=\gamma(R,t)\left(1-\beta(R,t)\mu\right)\alpha'_\nu(\nu')\\&=\gamma(R,t)\left(1-\beta(R,t)\mu\right)\gamma(R,t)^{-\frac{p+4}{2}}\left(1-\beta(R,t)\mu\right)^{-\frac{p+4}{2}}\alpha'_\nu(\nu),\\
    \alpha_\nu&=\gamma(R,t)^{-\frac{p+2}{2}}\left(1-\beta(R,t)\mu\right)^{-\frac{p+2}{2}}\alpha'_\nu(\nu). 
    \end{aligned}
\end{equation}

\subsection{Numeric flux integration}
We create a cylindrical coordinates grid of $N_r\times N_z$ ($N_r,N_z\sim$ few $1000$ each) cells, assuming azimuthal symmetry. The $z$ axis is aligned with the line of sight direction such that the radiation emitted by a cell at $t_\text{lab}$ arrives at the observer at
\begin{equation}
    t_\text{obs}=t_\text{lab}-\frac{z}{c},
\end{equation}
and
\begin{equation}
\begin{aligned}
    R &=\sqrt{z^2 + r^2},\\
    \mu &= \frac{z}{\sqrt{z^2 + r^2}}.
\end{aligned}
\end{equation}
The contribution to the observed intensity of each cell is given by
\begin{equation}
\begin{aligned}    
    \Delta I_\nu(r,t_\text{obs}) =& \frac{j_\nu\left(\sqrt{r^2+z^2},t_\text{obs}+\frac{z}{c}\right)}{\alpha_\nu\left(\sqrt{r^2+z^2},t_\text{obs}+\frac{z}{c}\right)}\left(1-e^{-\Delta \tau_\nu\left(\sqrt{r^2+z^2},t_\text{obs}+\frac{z}{c}\right)}\right)\times \\&e^{-\tau_\nu\left(t_\text{obs}+\frac{z}{c}\right)},
    \end{aligned}
\end{equation}
where $\tau_\nu$ is the optical depth between the cell and the observer and $\Delta \tau_\nu=\alpha_\nu \Delta z$. 


\bsp	
\label{lastpage}
\end{document}